

\input harvmac
\def \fourth { {1\over 4}}
\def \m {\mu}  \def \le { \l_{ \E }  }
\def \ce {c_{\rm eff}}
\def \ep {\epsilon}

\def \sn {\sqrt N }
\def \ra {\rightarrow}

\def \dvps {{\dot \varphi}^2}\def \dops {{\dot \phi}^2}
\def \dls {{\dot \lambda }^2}
\def \const {{\rm const} }

\def \dop  {\dot {\phi } } \def \dol {{\dot \lambda }}
\def \ddp {\ddot \phi }
\def \L {\Lambda}
\def \eq#1 {\eqno{(#1)}}
\def \e {\rm e}

\def \sf {string frame }
\def \ef {Einstein frame }

\def \ra {\rightarrow }
\def \e#1 {{\rm e}^{#1}}

\def \a {\alpha}

\def \sh {\ {\rm sinh} \ }
\def \ch {\ {\rm cosh} \ }
\def \th {\ {\rm tanh} \ }
\def \ln { {\rm ln } }
\def \sin {\ {\rm sin} \ }
\def \cos {\ {\rm cos} \ }
\def \l {\lambda}
\def \p {\phi}
\def \vp {\varphi}
\def  \g {\gamma}

\def \s {\sigma}
\def \t {\tau}
\def \b {\beta}

\def \pa {\partial}

\def \sqG {\sqrt {-G}}
\def \Goo {G_{00}}
\def \Gmn {G_{\mu \nu}}

\def \E {{\rm E }}
\def \CU {{\cal U } }

\def \CC {{\cal C} }
\def \CO {{\cal O}  }
\def \dl {\dot \lambda}
\def \ddl {\ddot \lambda}
\def \dvp {\dot \varphi}
\def \ddvp {\ddot \varphi}
\def \sm {\sum_{i=1}^{p}}

\def\np {  Nucl. Phys. }
\def \pl { Phys. Lett. }
\def \mpl { Mod. Phys. Lett. }
\def \prl {  Phys. Rev. Lett. }
\def \pr  { Phys. Rev. }

\Title{DAMTP - 15 - 1992
  \ \ hepth@xxx/9203033}
{\vbox{\centerline{ Cosmological solutions with dilaton }
\vskip2pt\centerline{ and maximally symmetric space in string theory}}}

\centerline{ A.A. Tseytlin
\footnote{$^\dagger$}
{e-mail: aat11@amtp.cam.ac.uk.   \ \
On leave of absence from the Department of Theoretical Physics, P. N.
Lebedev Physics Institute, Moscow 117924, Russia.}}
\bigskip
\centerline{\it DAMTP}
\centerline{\it Cambridge University}
\centerline{\it  Cambridge CB3 9EW }
\centerline{\it United Kingdom }
\baselineskip=20pt plus 2pt minus 2pt
\vskip .3in

We study time -- dependent solutions of the leading order string effective
equations for a  non-zero central charge deficit and curved maximally
symmetric space. Some regular solutions are found for the case
of non-trivial  antisymmetric tensor and vector backgrounds (in various
dimensions) and negative spatial curvature. It remains an  open question which
conformal theories  are exact generalisations of these solutions.
We discuss the analogy between the string cosmological solutions and
the solutions of the standard first order renormalisation group equations
interpolating between ``static" conformal theories.
\bigskip
\it{To appear in Int. J. Mod. Phys.:D      }
\Date{3/92} 


\newsec{ Introduction}

The standard approach to  string cosmology  is based on
the  analysis of time dependent solutions of the effective field theory
equations which may contain some ``matter" contributions (e.g. density and
pressure of a string gas ) but only the leading terms in the derivative ($\a'$)
expansion (see, for example, refs. [1-20]).  This approach is justified under
the
assumption that  the evolution is ``adiabatic" (i.e. that gradients are
small). This assumption  may be valid at late enough times (e.g. after the
supersymmetry is broken and dilaton got a mass).  However, in order to describe
an early (``string") phase of evolution and, in particular, to understand
the issue of cosmological singularity  it is necessary to go beyond the
derivative expansion.  There is no much sense in including some  of higher
derivative terms  (e.g. $R^2$ ones ) in the effective action since if $\a'$
corrections $are$ important one should account for all of them on an equal
footing.

The only known alternative to the $\a'$- expansion is to find an exact 2d
conformal theory which has an  appropriate  ``cosmological" interpretation.
While the first attempt in this direction used a direct product of the
``time"  and ``3-space" conformal theories [5] some  non-trivial
time-dependent  conformal theories based non-compact coset models associated
with gauged Wess -Zumino - Witten theories  were recently considered in
[21--27] (see also [28]). This approach has its own obvious limitations. A
conformal theory corresponds only to a perturbative (classical) solution of a
superstring (Bose string) theory. It is not known whether non-perturbative
solutions (e.g. extremals of an effective action which contains
non-perturbative
corrections like a  dilaton potential) can be described in terms of 2d
conformal
theories. One may hope that the two approaches -- the ``phenomenological" one
(based on a low energy effective action containing only leading terms in the
$\a'$ expansion but including non-perturbative corrections and ``matter" terms)
and the ``conformal" one may be complementary in the sense that they apply at
different scales (times). For example, if at early enough times  the dilaton
potential is not yet generated, i.e. the dilaton is massless,   one may hope to
describe the cosmological evolution by an exact conformal field theory.

The usual method of
giving a space-time interpretation to a conformal theory is based on relating
it
(in some approximation) to a solution of the leading order string effective
equations (Weyl invariance conditions for the corresponding $\s$ model). While
in $D=2$ the most general solution [9,29] has the known conformal
field theory counterpart [22] (see also [17]) the situation in higher
dimensions $D \geq 3$ is very different: for most of the solutions of the
leading order effective equations  their conformal field theory generalizations
are unknown. Conformal theories based on gauged WZW models  seem to
correspond to a rather small and special  subclass of  solutions of the $D\geq
3$ effective equations. In particular, they often have singularities
and very few if any symmetries [23--28].  For example, only a subset of the
simplest ``toroidal" cosmological solutions with $N=D-1$ commuting isometries
[9] has known  conformal field theory realisation [27].

The standard cosmological backgrounds   have their spatial parts
represented by   maximally symmetric $N$ - dimensional manifolds (or, in the
Kaluza-Klein context,  products of maximally symmetric manifolds), i.e. a
sphere,
a flat space or a pseudosphere.\foot{One should keep in mind, of course, that
the assumption of maximal symmetry may not be a valid one at early
times (e.g. before an inflationary stage).} Since there should be such regular
solutions of the effective string equations one is naturally led to the
question about their possible conformal field theory generalisations. It is
striking that the only example of a  solution with maximally symmetric space
which has  known   conformal field theory interpretation
is  the  ``static" $N=3$ solution [5]  corresponding to
the direct product of  the $D=1$ (``time") conformal theory with linear
dilaton [4] and $SU(2)$ WZW model [30]. The $N=3$ (pseudo)sphere is special
being equivalent to a group space. $N\not= 3$ spheres and  $D\not= 3$  de
Sitter
spaces do not directly correspond to conformal theories. For example,
the ``(anti) de Sitter string" of ref.[21] based on  gauged  $SO(D,1)/SO(D-1,1)
$  $\ ( SO(D,1)/SO(D-1,1) ) \  $ WZW model with $D>26 \  $ $ \ ( D<26) \
 $ has a space-time interpretation not in terms of the de Sitter space
but a  background  which does not have a maximally symmetric space [24--26].

One may hope to grasp some features of  hypothetic ``maximally symmetric"
conformal theories by studing  properties of particular solutions of the
leading
order string effective equations. This is one of the motivations behind the
analysis of  cosmological solutions in the present paper.
Some of the solutions with maximal spatial symmetry may play a  fundamental
role, e.g.  as building blocks for more complex cosmological backgrounds. In
particular, it seems important to generalize the solution of ref.[9] (its
isotropic  limit) which has flat space sections to the case when the space has
a
non-zero curvature.
 That is why we shall not restrict the number of spatial dimensions $D$ and the
value of the central charge deficit $c$.

Another motivation for a study of the time - dependent solutions
is a possible application to a description of a
``late time" string cosmology within the ``phenomenological" approach mentioned
above. Since the precise form of  non-perturbative corrections to the effective
action (e.g. to the dilaton potential) is unknown one would like
to understand the behaviour of the solutions  in various special cases in order
to extract their universal features, in particular, some generalities in the
evolution of the dilaton (which is non-trivial  except  in   the radiation
dominated phase  where a $\p = \const$ solution may exist [17,13,20]).

Most of the previous discussions of the string theory cosmological solutions
(see e.g. [6,7,19,20]) used the
``Einstein frame", i.e.  the parametrisation of the metric in which there is
no dilaton factor in front of the Einstein  term in the effective action.
Though the \ef [31] and  the \sf [32]  should be physically equivalent, the \sf
metric  (i.e.  the ``$\s$ model metric" [32]) is the one  strings directly
interact with. That is why it is natural to use the \sf in
order to try to understand  a string theory interpretation of the solutions
 (see also [10,11])  and,
in particular, to trace a possible  relation to exact  conformal  field
theories.

In Sect.2 we shall review the general structure of the cosmological evolution
equations in arbitrary dimension $D$. The use of the \sf and ``shifted" dilaton
field [14,15] simplifies their form and makes a qualitative analysis of the
solutions particularly transparent.

In Sect.3  we shall consider the cosmological solutions in the case of
non-trivial antisymmetric tensor  and vector backgrounds.\foot {
The presence of a non-zero central charge deficit $c$  corresponds in the
Einstein frame to  the  exponential dilaton potential. Similar systems  were
considered  in the context of Kaluza-Klein cosmology  based on higher
dimensional supergravity models  and in connection
with $D=4$ superstring theory (see e.g. [33--35,3] and [7,19,20]).}
 We shall see that for $c<0$ the decreasing dilaton produces a damping force
which stabilizes the solutions. We shall find  a new  analytic
solution in the case of a rank $N$ antisymmetric tensor background and zero
spatial curvature.
The  asymptotics of this solution is the same as that of
the isotropic case of the vacuum solution of ref.[9], i.e. for $t \ra \infty$
the scale factor approaches a constant while the dilaton is linearly
decreasing.
This does $not$ necessarily imply that  conformal theories which should
correspond to these solutions are  also the same and  are trivial, i.e.
equivalent to the direct product of time  and  flat $N$- space theories.\foot
{ For example,  the solutions in refs. [22--28] and their asymptotics
correspond to different conformal theories. }

In Sect.4 we  shall study solutions with negative spatial
curvature and $c<0$. Solutions with $c>0$  (or positive curvature in
the absence of ``matter") appear to be singular.  In particular, we shall find
a
regular  vacuum solution  (for arbitrary $D$) which is probably the ``closest"
string theory analog of the open de Sitter space.\foot{ Our solution is
different from the $D=4$ solution of ref.[19] (even though they  have the
same large time asymptotics in the \ef). The latter is a  special ``power law"
 solution which exists only in the presence of the $R^2$ term in the
effective action.}  Though at small times it resembles the $D$
-- dimensional de Sitter space [17] its large $t$ behaviour is different:
the scale factor is expanding as a power of time and not exponentially. The
difference from the de Sitter space  can be attributed  to the
non-trivial time dependence of the dilaton field. If the spatial curvature is
non-zero there are no solutions with a constant dilaton even for $c=0$.

The large $t$ asymptotics  of the  solutions of Sect.4 in the \sf is different
from that of the solutions of Sect.3 and of refs.[4,5,9]: the scale factor is
$not$
approaching a finite constant, i.e. asymptotically the space--time is not  a
trivial product of time and space factors. This suggests that their
 conformal field theory generalisations should be  different as well.
 Let us note that  being
transformed to the \ef all the solutions (with $c \not= 0$) we have discussed
have the same large time behaviour: the linear growth of the scale factor and
the
logarithmic decrease of the dilaton. This illustrates the point that it is not
sufficient to find an asymptotic behaviour of a solution in the \ef in
order to identify  the corresponding  conformal theory.

The relation between cosmological solutions in
the \sf and the \ef will be studied in Appendix A. We shall find how the
asymptotics of our solutions look like in the \ef.

In Appendix B we shall demonstrate  (using the analytic
continuation which interchanges space and time coordinates)  the correspondence
between the $N=1$ case of the solution found in Sect.3   and the  charged black
hole solution of $D=2$  heterotic string theory [36].

In Appendix C we shall consider time--dependent solutions in the presence of an
extra scalar field (e.g. a modulus of an extra compact dimension or a coupling
corresponding to a nearly marginal perturbation of a conformal theory). We
shall
discuss a relation between string
equations (which are of second order in ${d \over d t}$) and the standard
renormalisation group equation (which is of first order ${d \over d t}$)
 providing two
alternative pictures of an  interpolation between different conformal points
(cf. [39--41]). This illustrates the point that  asymptotics of cosmological
solutions may be related  to different conformal theories (as well as being
different from  a conformal theory which
generalizes the time--dependent solution itself). While solutions of the
standard RG equations interpolating between ``static" (or ``N - dimensional")
conformal theories are
not related to any ``$N+1$ - dimensional" conformal theory, solutions of string
equations
 should be interpretable as  ``$N+1$ - dimensional" conformal theories.

  \newsec{ Cosmological equations for the scale factor and the
dilaton }

  Our basic assumptions will be the following:  (i) ``adiabaticity" of
cosmological evolution (in particular, a possibility to ignore higher
derivative
terms in the effective action); (ii) weak coupling (we shall ignore string
loop corrections and impose the condition that the dilaton or the effective
string coupling should not increase with time); (iii) maximal spatial
symmetry. We shall study the following FRW-type cosmological background
$$ds^2 = - dt^2 +  a^2 (t) \ d\Omega^2 \ \ , \ \ \  d\Omega^2 =  g_{bc}dx^b
dx^c\  \  , \eq{2.1} $$  $$ a(t) = {\rm e}^{\l (t)} \ , \ \ \p=\p(t) \ ,  \ \
b,c= 1, ..., N \ , \ \  D=1 + N \ \ , $$
where $ g_{bc}$ is the metric of a maximally
symmetric $N$ dimensional space  with the radius of curvature $k^{-1} \ \
(k=-1,0,1) $, i.e. $R_{bc}= k(N-1) g_{bc}$. \foot {One may, of course,
consider a Kaluza--Klein --type cosmological ansatz with  the space being a
product (with different time dependent radii) of several maximally symmetric
factors (see e.g. [33-35] and eqs. (2.24)--(2.31) below).}

  The metric we shall consider corresponds to
  the ``string frame"  in which the leading
order terms in the low energy expansion of the effective action of a
closed string theory  have the  following form (we  absorb the gravitational
coupling constant into $\p$ and   set $\a'$=1)
$$ S =  \int d^D x \sqG \  \e{-2\p}   \ [ \ c + \ R \ + 4 (\pa \p )^2
  -V(\p) - L_m\ ] \ . \ \eq{2.2} $$
Compared to the corresponding equations in the Einstein frame the resulting
cosmological equations have  simpler structure  and more natural interpretation
from string theory point of view.
 We shall discuss the relation between
cosmological solutions in the string frame and the Einstein frame in Appendix
A.   $V$  in (2.2) is a dilaton
potential and  $$ c= - {2\over 3\a'} (D_{\rm { eff}}- D_{{\rm crit}})  \
$$ depends on details of  particular string
theory ( $D_{\rm { eff}}= D \ , \ D_{{\rm crit}} =26 $ in the Bose string
theory,  $D_{\rm { eff}}= {3\over 2} D \ , \ D_{{\rm crit}} = 15 $ in the
superstring theory , etc). We
consider $c$ as an arbitrary parameter and prefer not to include it into $V$
(for  discussions of ``non-critical" string cosmology see e.g.
[4,5,9,17,18]). The ``matter" Lagrangian contains contributions of other
``light" degrees of freedom (for example, vectors and the antisymmetric tensor
$B_{\mu \nu}$). We can represent the sum of each of the matter field kinetic
terms in the following  generic way $$ L_m= \sum_{n=1,2,3} {1 \over 2n!}
H^2_{\l_1  ... \l_n} + ... \ , \ \
 H_{\l_1  ... \l_n}= n \pa_{[ \l_1} B_{\l_2 ... \l_n ]} \ \ . \ \eq{2.3} $$
The combinations of the  metric and dilaton equations which follow from (2.2)
are
$$ R_{\mu \nu} + 2 D_{\mu} D_{\nu} \p = - {\fourth } {\pa V \over \pa \p}
G_{\mu \nu}  +
 { \e{2\p}  \over \sqG  } ({\delta S_m \over \delta G^{\mu \nu}} -
 \fourth  G_{\mu \nu} {\delta S_m \over \delta \p}) \ \ , \eq{2.4} $$
 $$ c +2 D^2 \p - 4 (\pa \p)^2 = V + \fourth (D-2) {\pa V \over \pa \p}
 +  { \e{2\p}  \over \sqG  }
 [ \ \fourth (D-2){\delta S_m \over \delta \p}  -
 G^{\mu \nu} {\delta S_m \over \delta G^{\mu \nu}} \ ] \ \ . \ \eq {2.5} $$
 In the case of the metric (2.1) one finds from (2.4), (2.5)
 $$ \ddl +N\dls - 2 \dop \dl + k (N-1) \e{-2\l} =
 - \fourth {\pa V \over \pa \p} - {1 \over 2N } \e{2\phi - N\l}
 ({\delta S_m \over \delta \l} + \ha N {\delta S_m \over \delta \p} ) \ \ ,
 \eq {2.6} $$
 $$ 2 \ddp - N \ddl - N \dls = \fourth {\pa V \over \pa \p}
 - \e{2\phi - N\l } ({\delta S_m \over \delta \Goo } -
 \fourth {\delta S_m \over \delta \p} )
 \ \ ,  \eq {2.7} $$
 $$ c + N(N-1) \dls + 4 \dop^2 - 4 \dop \dl + k N(N-1) \e{-2\l } =
 V  -2 {\e{2\phi - N\l } } {\delta S_m \over \delta \Goo }\ \ , \eq {2.8} $$
 where $\Goo = -1 $ after the variation.  As is well
known, eq. (2.8) is the
 ``zero energy" constraint which is conserved as a consequence of (2.6) and
(2.7) and hence gives only a restriction on the initial values of
$\dop , \dol , \p , \l $. The structure
of the system (2.6)--(2.8) is made more transparent by introducing the
``shifted"
dilaton field  $\vp$ [ 14,15,17]
$$ \vp \equiv 2\p - N \l \
\ ,\  \ \ \sqG\   \e{-2\p} =  {\sqrt g }\ \e{-\vp} \  \ . \eq{2.9} $$
Then  (2.2),(2.6)--(2.8) take the following form (we absorb the constant factor
of the volume of $N$--space into $\vp$ )
$$ S=\int dt\  \e{-\vp} \sqrt
{-\Goo}\  [ \ c - G^{00} N \dl^2
 + G^{00}\dvp^2 \ ]  - S'_m [G_{00}\ , \l \ , \vp] \ \ , \eq{2.10} $$
$$ c- N \dl^2 +\dvp^2 = 2U \ ,
\eq{2.11} $$
$$\ddl - \dvp \dl = W_1   \ \ , \ \ \  \ \ddvp - N\dl^2 = W_2 \ \ , \
\eq{2.12} $$
where
$$ U = - \ha  k N(N-1) \e{-2\l} + \ha V
- \e{\vp} \ {\delta S_m \over \delta G_{00}} \ \ , \eq{2.13} $$
$$ W_1 = - k (N-1) \e{-2\l} -\ha {\pa V \over \pa \vp}
 - {1 \over 2N }\e{\vp} \ {\delta S'_m \over \delta \l}\ \ , \eq {2.14} $$
$$ W_2 = \ha {\pa V \over \pa \vp}-
{\e{\vp} }( \ {\delta S'_m \over \delta G_{00} }  - \ha {\delta S'_m \over
\delta \vp} \ )  \ \ , \eq {2.15} $$
and
$$ V(\l\ , \vp ) = V(\p) \ \  , \ \ \ S'_m [G_{00}\ , \l \ , \vp]=
S_m [G_{00}\ , \l \ , \p] \ \ , \ \ \p= \ha ( \vp + N \l) \ \ . \eq{2.16} $$
The conservation of (2.10) implies ( we are assuming
of course that $S_m$ is covariant under reparametrisations of time)
$$ { \pa U \over \pa \l} = - N W_1 \ \ ,
\ \  { \pa U \over \pa \vp} = W_2 \ \ . \eq{2.17} $$
The resulting system
$$ c- N\dls + \dvps = 2U  \ \ , \eq{2.18} $$
$$  \ddl - \dvp \dl = - {1 \over  N} {\pa U \over \pa \l} \ \ , \eq{2.19} $$
$$  \ddvp - N \dls = {\pa U \over \pa \vp }  \ \ ,  \eq{2.20} $$
has an obvious mechanical interpretation and is very useful for a qualitative
analysis of the solutions  [17,18].
Note that eq.(2.19) and  a linear combination of (2.20) and (2.18)
can be derived from the action (cf. (2.10))
$$ S= \int dt\  \e{-\vp} [ \ c + N\dl^2
 -\dvp^2 \   - 2U ( \l \ , \vp) \ ] \ \ . \eq{2.21} $$

Assuming that $\dl >0$ (expansion)  $\ \dvp <0$ is a necessary condition for
a decreasing of the dilaton, i.e. $\dop <0$.  Suppose ${\pa U \over \pa \vp }
\ge 0$. Then  (2.20) implies that $\ddvp >0$. If also $2U -c \ge 0$
we get $\dvp \not= 0  $ from (2.18).  As a result, if  the initial value
$ \dvp (0) $ is negative  (as we shall always assume)
 $ \dvp$ will never change sign  and so $\vp$ will always
decrease. Then eq.(2.19) describes  a motion of a particle in a
potential (in general time--dependent through $\vp (t)$ ) with a damping term
$\sim \dvp$.  Because of the  dilaton damping effect the ``energy"
$$ E= \ha N \dls + U(\l,\vp) = \ha ( c + \dvps)  \ \eq{2.22} $$
is decreasing with time
$$ {dE\over dt} = \dvp ( N\dls + {\pa U \over \pa \vp } ) < 0 \ \ . \eq{2.23}
$$

One can generalize the above cosmological equations to the case
of a Kaluza--Klein --type cosmological ansatz taking
the space to be a product of
several $(i=1,...,p)$ maximally symmetric factors of dimensions $N_i$
 $$ ds^2 = - dt^2 + \sm a_i^2 (t) g^{(i)}_{ b_ic_i}dx^{b_i}_i
dx^{c_i}_i \  \  ,\eq{2.24} $$
$$ a_i= {\rm e}^{\l_i (t)} \ , \ \ \p=\p(t) \ ,
\ \   \  b_i , c_i = 1,...,N_i   \ , \ \  N=\sm N_i\ , \  D=1 +
N \ \ , \eq{2.25} $$
$$R^{(i)}_{b_i c_i} = (N_i-1) k_i g^{(i)}_{b_i c_i}\  \ \
\ . \eq{2.26} $$
Introducing the ``shifted" dilaton
$$ \vp = 2\p - \sm N_i \l_i \ \ , \eq{2.27} $$
one finds that the system  (2.18)--(2.20) is replaced by (cf.
(2.13)--(2.16))
 $$  c - \sm  N_i\dl_{i}^2  + \dvp^2  = 2 U   \ \ , \ \eq{2.28}  $$
$$ \ddl_i - \dvp \dl_i  =   - {1\over N_i} {\pa U \over \pa \l_i}\ \ ,
\eq{2.29}  $$
$$ \ddvp_i - \sm  N_i\dl_{i}^2 =  {\pa U \over \pa \vp}   \ \ , \eq{2.30} $$
$$ U = - \ha \sm N_i (N_i-1) \ k_i \e{-2\l_i} + \ha V
 -\e{\vp} {\delta S_m \over \delta G_{00}} \ . \eq{2.31} $$

\newsec {Solutions with  antisymmetric tensor
backgrounds}

1. We would like to understand the behaviour of the solutions of the system
(2.18)--(2.20)  for various choices of the ``potential" $U(\l, \ \vp)$.
The simplest possibility  when the ``matter" and dilaton potential are absent
but the space has a non-zero curvature will be
discussed in the next section. Here we consider the case when the ``matter"
part of the  action (2.2) contains only the  antisymmetric tensor terms
(2.3).\foot{ Cosmological solutions  with antisymmetric tensor backgrounds
were studied in Kaluza--Klein context [33-35]. For a discussion of
cosmological solutions of a similar system with dilaton in the \ef see also
 [3,7].}
The equation for the   antisymmetric tensor of rank $n-1$ is  $$
D_{\l_1 } ( \e{-2\p} H^{\l_1 ... \l_n} ) = 0 \ \ . \eq{3.1} $$  A
non-trivial solution consistent with symmetries of the ansatz (2.1) is found if
the number of space dimensions $N$ is equal to the rank  $n$ of the
antisymmetric tensor field strength [33] $$ H_{ 0
a_1...a_{N-1} } =0 \ , \ \  H_{  a_1...a_{N} } = h \ep_{  a_1...a_{N} } \ , \ \
h=\const \  \ , \eq{3.2} $$ where $\ep_{  a_1...a_{N} }$ is the standard
covariantly constant antisymmetric tensor. Then $U$ in (2.13) takes the form
$$U =- \ha  k N(N-1) \e{-2\l}  + {1 \over 4N!} H^2_{\l_1  ... \l_N}
+ {1 \over 2(N-1)!} H^0_{a_1  ... a_{N-1}} H^{0 a_1  ... a_{N-1}}  \eq{3.3} $$
$$ = - \ha  k N(N-1) \e{-2\l} + \fourth h^2 \e{-2 N \l} \ \ .  $$
The two particular  well known cases correspond to $N=2 $ (``monopole"
background of a vector field [35]) and $N=3$ [33,34]
$$  N=n=2 \ , \ \  F_{ab}=h \ep_{ab} \ , \ \
U= -k\e{-2\l} + \fourth h^2 \e{-4 \l}  \ , \eq{3.4} $$
$$  N=n=3 \ , \ \ H_{abc} = h \ep_{abc} \ , \ \
U= -3k\e{-2\l} + \fourth h^2 \e{-6 \l}  \ . \eq{3.5} $$
If the space curvature is $positive$  $k>0$  (i.e. the space is
a $N$-sphere)
the potential $U$ has the minimum at $\l=\l_0$
$$ h^2 = 2k (N-1) \e{2(N-1) \l_0} \ \ , \ \ \ \
 U(\l_0) = - \fourth (N-1) h^2 \e{-2 N \l_0} \ < \ 0 \ \ . \eq{3.6} $$
If $c<0$ the system (2.18)--(2.20) has the following  ``static" solution
$$ \l=\l_0 \ , \ \ \ \  \vp=\vp_0 - 2bt \ \ ,  \ \ \ \  \p =\p_0 - bt \ \ ,
 \eq{3.7} $$
$$ 4b^2 = -(c- 2U (\l_0) )  = |c| - \ha (N-1) h^2 \e{-2 N \l_0} \ . \eq{3.8}
$$ We have assumed that $\l_0$ (or $h$) and $c$ are such that $b^2 \geq 0$.
The corresponding solution in the \ef is given in (A.16).
For $N=3$ this is the solution found in
[5]. It has the obvious  conformal field theory generalisation [5]
represented by the direct product of the $D=1$ ``time" theory with linear
dilaton
[4] and the $SU(2)$  WZW theory (i.e. $S^3$ parallelised by the antisymmetric
tensor background).  The $N=2$ case (with $b=0$) was considered  in [35].\foot{
In this case the
constant $F_{ab}$--flux ``compensates"  for the  curvature of
$S^2$. It is possible to interpret the $SU(2)$ WZW model as an exact conformal
field theory which generalizes this solution to all orders in $\a'$ expansion
 [37].}

 As we discussed in sect.2, if $c<0 \ \ $  $\dvp $ remains negative if it was
negative at $t=0$. That means the dilaton term in (2.19) plays the role of a
damping force. As a result, the solution (3.7) is $stable$  so that for a broad
range of initial conditions a
 time - dependent solution with decreasing dilaton approaches (3.7), i.e. the
space-time is asymptotically $R \times S^N$
(since the ``energy" (2.22) is decreasing with time the trajectory on the
$(E,\l)$ plane is approaching the minumum of $U$, reflecting from the walls of
$U$ ).\foot {In Kaluza--Klein context these solutions  provide a
model of ``confinement" of internal dimensions (because of the damping by
dilaton there are no oscillations of internal radii, so that the effective
couplings also do not fluctuate). }

The solution (3.7) exists  only if
$$ c_{\rm eff} \equiv c -2U(\l_0) \leq 0    \  \ . \eq{3.9} $$
If $\ce >0$ the sign of $\dvp$
changes at the point where $ 2U = c$. After this happens the dilaton term in
(2.19) provides an accelerating force and the solution goes to infinity in a
finite time. In general, it appears that all solutions with $\ce >0$ (or $c >0$
if $U$ is non-negative and  approaches zero at large $\l$ )
are singular.\foot{ In the bosonic
string $c>0$ corresponds to $D<26$. Singularity of solutions with $c>0$
was already observed in [9] in the simplest case of a flat space  ($k=0$)
and the absence of matter.}
Since according to (2.22) $E \geq \ha c $  the trajectory $E(\l)$  with the
initial condition $\dvp < 0$  first goes down, reflects off the line $E=\ha c$
and then goes  up to the infinity in a finite time.

If $k \leq 0$, i.e. the curvature of the space is negative or zero,  $U$
in (3.3) is positive and has no local minima. A qualitative behaviour of
solutions  does not significantly depend on the presence  or absence of the
antisymmetric tensor $O(h^2)$ term in  $U$. We shall study the case of
$k<0 $ in detail in the next section.

2. Let us now consider another class of cosmological solutions
generated by  the following solution of (3.1) [33]
which exists if
$n=N+1$ \foot{ Since in a closed string theory context the rank $n \leq 3$ this
solution is possible  only in $D=2,3$. However, one can  consider a product of
the corresponding  $D=2$ or $D=3$ space - time and additional space factors
thus enlarging the total dimension.}
$$H_{\l_1 ... \l_{N+1}}= h \ \e{2\p } \ep_{\l_1 ... \l_{N+1}} \ \ , \ \
h=\const
\ \ . \eq{3.10} $$
Then
$$ H_{\l_1 ... \l_{N+1}}^2 =- (N+1)!\  h^2  \e{4\p} \ ,
\eq{3.11} $$
so that the potential $U$ in (2.13), (3.3) takes the form
$$ U = -\ha  k N(N-1) \e{-2\l} + \fourth h^2 \e{ 4\p} =
  -\ha  k N(N-1) \e{-2\l} + \fourth h^2 \e{ 2\vp + 2N\l} \ \ . \eq{3.12} $$
The antisymmetric tensor contribution to $U$ in this case is the same as
that of the ``two-loop" term in the dilaton potential (cf. (2.13))
$$V= \ha h^2 \e{ 4\p} \ \ . \eq{3.13} $$
Let us first ignore the effect of the space curvature and set
$k=0$ (we shall consider  solutions with non-zero $k$ in the next section).
Remarkably enough, the  system (2.18)--(2.20) corresponding to the case of
$k=0$
and $V$ given by (3.13), i.e.
$$ c-N\dls + \dvps = 2U  \ \ , \eq{3.14} $$
$$  \ddl - \dvp \dl = -2U\ \ ,\eq{3.15} $$
$$  \ddvp - N \dls = 2U \ \  \eq{3.16} $$
has a simple analytic solution. In fact, subtracting (3.14) from (3.15) we get
$$ \ddot y + cy = 0 \ \ , \ \  y \equiv \e{-\vp} \ \ . \eq{3.17} $$
To have a regular solution with $\vp$ decreasing with time we need to assume
that $c \leq 0$. Then (for $y(0)=0$)
$$ \vp = \vp_0 - {\ln  \sh 2} bt   \ \ ,\ \  4b^2 = -c \ . \eq{3.18} $$
This is the same $\vp$ as in the solution of (3.14)--(3.15) in the case of
$U=0$
(the isotropic case of the solution of ref.[9])
$$\vp = \vp_0 - {\ln  \sh 2} bt   \ \ , \ \
  \l = \l_0  \pm {1\over \sn}\  {\ln \th }bt \ \ . \eq{3.19} $$
The solution for $\l$ is found by substituting
(3.18) into (3.14) and integrating the resulting equation for
${\rm e}^{-N\l }$
$$ \l = \l_0  - { 1 \over N}\  {\ln \ } [\  A^{-1}{({\th
}bt)}^{-\sn} + A \ {({\th }bt)}^{\sn} \ ]\ , \eq{3.20} $$
$$ \p = \p_0 - \  \ha \ln \ ( \sh bt \ch bt \ [\  A^{-1}{({\th
}bt)}^{-\sn} + A \ {({\th }bt)}^{\sn} \ ]\ ) \ , $$
$$ A^2 = {h^2 \over 32
b^2} \e{2\vp_0 + N \l_0 } = {h^2 \over 8 |c|} \e{4\p_0} \ \ . $$
The large $t$ behaviour of the solution (3.20), (3.18) is
$$ \l \ra \l_0 - {1\over N  } \ \ln \ (A+A^{-1})  = {\const } \ \ , \ \
\vp = \vp_0 - 2b t \ \ , \ \ \p = \p_0 - bt  \ .  \eq{3.21} $$
In the special case of $c=0$ one finds
$$\vp = \vp_0 - {\ln \ }t   \ \ , \ \ \l = \l_0  - { 1 \over N}\  {\ln}\
(A^{-1}t^{-\sn}+ A t^{\sn}) \ , \eq{3.22} $$
so that the scale factor  $a$ grows at small $t$ until it reaches its maximum
at
$t_*= A^{- 1/ \sn  } $ and then asymptotically contracts to zero.The
dilaton $\p$ first grows and then starts decreasing (see also Appendix A).

If $h=0$ ( i.e. if  $U=0$ )  (3.20)
reduces to (3.19). $\pm$ in (3.19) indicates the two solutions related by the
duality  transformation [14,15]. The solution (3.18),(3.20) is invariant under
the shift $bt \ra bt + \ha i \pi $ , i.e. ${\th }bt \ra ({\rm tanh \ }bt)^{-1}\
,
\ $ etc  (which relates the two solutions in (3.19))  combined with $A \ra
A^{-1}$. Thus the system (3.14)--(3.16) with $h^2 /|c|
\ra  ( h^2 /|c| ) ^{-1} ,
\p_0 \ra - \p_0 $ (i.e. $A \ra A^{-1}$)  will have the ``dual" solution
(3.20) with ${\th }bt \ra ({\rm tanh \  }bt)^{-1}$.  This
transformation  is reminiscent of a ``weak coupling--strong coupling" duality.

Since ${\rm e}^{ \vp }$ is decreasing with time the same is true for the second
term in the  potential $U$ in (3.12). Therefore it is not surprising that the
behaviour of  (3.20) is similar to that of (3.19) : the two solutions  look the
same  for small $t$, then $\l$ grows (and, if $A >1$,   reaches a  maximum at
$t_*$ , $ A^2 \ {({\th }bt_*)}^{\sn}=1$) and finally approaches
the constant  value  (3.21)   at $t\ra \infty$.
We shall show in Appendix A that in the Einstein frame
such behaviour corresponds to a linear expansion of the scale factor
 (note that asymptotically the dilaton (3.18) is linearly
decreasing with time, $\vp \ra \vp_0 - 2bt$). As was noted in [14] the
analytic continuation  ($t \ra ir \ $ and $ \  c\ra -c \  \ $ or $ b\ra -ib$)
of the $N=1$ vacuum  solution (3.19)  coincides with the (euclidean)  ``black
hole" solution of $D=2$ Bose string theory [29,22].  In Appendix B we shall
demonstrate that the analytic continuation  of the $N=1$ case of (3.20) (which
interchanges the space and time coordinates and  $\  c\ra -c \ , \ $  i.e. $
b\ra -ib\ , \  A\ra - A $ ) is equivalent to  the ``charged black hole"
solution
of $D=2$ heterotic string theory found in [36].

\newsec {Solutions with negative spatial curvature}

In the absence of ``matter" and dilaton potential the system (2.18)--(2.20)
takes the following form
$$ c- N\dls + \dvps = -N (N-1) k \ \e {-2\l} \ \ , \ \ \  \eq{4.1} $$
$$  \ddl - \dvp \dl = - (N-1) k \ \e {-2\l}\ \ , \eq{4.2} $$
$$  \ddvp - N \dls = 0 \ \ .  \eq{4.3} $$
One may ask how ``close" can string solutions resemble
 the  maximally symmetric $D$ - dimensional  de Sitter space.\foot{ The
question
 about possible relation between solutions of (4.1)--(4.3) and ``de Sitter"
coset
conformal field theories [21]  was raised in [17] and also studied in [38]. As
we
have mentioned in the introduction, no direct connection seems to exist.}
Naively, one may expect that the role of $c$ in (4.1)  is similar to that of
the cosmological constant in the corresponding Einstein equation.  In fact,
rewriting (4.1) in terms of the original dilaton $\p$  we get
$$   N(N-1)\dls +
4\dops -4N \dop \dl = \L - N (N-1) k \e {-2\l} \ , \ \ \ \L \equiv -c \ \ .
\eq{4.1'} $$ If $\p=\const \ (4.1') $ has the following solutions:
$$  c <0 \ ({\rm de\  Sitter })  : \  \e{\l } = H^{-1 }\ch Ht  \ \ \ (k=+1) \ ;
$$  $$\  \e{\l } = H^{-1 }\sh Ht \ \  \ (k=-1) \ ; \ \ \l = \l_0 + Ht \ \
(k=0) \ \ , \ \ H^2=   -{c \over N(N-1) }\ \ , \eq{4.4}   $$
$$ c>0 \ ( {\rm anti\ de \ Sitter }) : \ \e{\l } = H^{-1 }\sin Ht  \ \
\ (k=-1) \ , \  \ H^2= {c \over N(N-1) }\ \ . $$
The point, however, is that while $\p=\const $  and (anti) de Sitter metric
solve (4.1) and (4.2) they do not satisfy the remaining dilaton equation
(4.3).\foot {For a similar observation in the \ef see Boulware and Deser [1]
and
also [6].} That is why the dilaton should necessarily change with time  and
that
produces
 a ``deformation" of the de Sitter metric. In fact, it turns out that
in the  asymptotic region of  large $t$ it is the time variation of the dilaton
and not that of the scale factor that ``compensates" for the presence of the
``cosmological constant" $-c$  in $(4.1')$.

  Let us consider the most interesting case when $c\leq 0  \ $ and
the space has a negative curvature  $ \ k < 0 $ (if
$c$ or $k$ are positive $\dvp$ may change sign at some point and the solution
is
singular, see also below). Then $c_{\rm eff}=c + N (N-1) k {\rm e }^{-2\l }$
in (3.9) is negative and hence if $\dvp (0) <0 $ the
dilaton $\vp$ is always decreasing with time  (note that (4.1)
is a constraint on initial values of $\dl \ , \dvp \ , \l  $ so that
$\dvp (0) \leq - \sqrt { |c_{\rm eff}| } $ ). The potential  $$U=- \ha N (N-1)\
k\ \e {-2\l} \   $$ grows at negative $\l$  and thus prevents penetration into
the region of small scales $a={\rm e}^{\l }$. If $a$ is contracting  at the
initial moment it  reflects from the potential wall and eventually expands to
infinity.\foot {The existence of a minimal radius of contraction in the case of
the negative spatial curvature is similar  to what was found in [17,18] in the
case of cosmological evolution in the presence of the classical fields
corresponding to the ``momentum" modes of a string compactified on a torus. In
fact, the (mass$)^2$ of momentum modes is proportional to $ a^{-2}$  and hence
the resulting cosmological system is similar to (4.1)--(4.3). The conclusion
was [17]  that the contribution of the  classical momentum modes prevents
the scale factor from  contracting to zero. }
The trajectory of the system on the energy plane $E(\l)$ is going down
because of the damping effect of the dilaton (see (2.22), (2.23)).

Let us assume that $\l (0) $ is  very large and negative
(i.e. the scale factor $a$ is very small). To study  solutions
at small times we may  drop $c$ in (4.1) since the potential term dominates.
Then
the system (4.1)--(4.3) has the following  special solution
\foot { For $c=0$ one can solve (4.1)--(4.3) analytically. Introducing
$f= \dvp$ one finds:
$ \ f'={df \over d\l } = \sn \sqrt { f^2 - m^2 \e{-2\l } } \ ,
\ m^2= N (N-1)|k |
$.   This gives $ w' = w +\sn \sqrt { w^2 - m^2 }  \ , \  w= \e{\l } f $
 which is easily integrated.}
$$ \l=\l_0 +  \ln \ t \  ,  \ \ \vp = \vp_0  - N\ \ln \ t \  , \ \ \p=\p_0
={\const } \ , \ \ - k\ \e{-2\l_0 } =1 \ . \eq{4.5} $$
If $N > 1$ such a solution exists only for the $negative$ spatial curvature
$k<0$
[17]. The corresponding energy $E$ is infinite at $t=0$ so that the expansion
starts from $\l=- \infty $.  This solution describes a flat space (the scalar
curvature is $ R = -2 (\ddl +\dls ) - N ( \dls + k \e{-2\l } ) =0 )$  and
may be interpreted  (at small $t$ )  as a $D=N+1$ dimensional open de Sitter
space (4.4) ``born" at $t=0$ (note that the dilaton $\p$ is approximately
constant at early times).

Solutions with regular initial conditions (finite $\dvp (0) \ , \ \dl (0) $ )
correspond to an expansion from some minimal non-vanishing $a$.  One could
expect
that  the large $t$  asymptotics of the solution  should coincide with that
of the $k=0$ solution (3.19)  because the effect of the potential  should be
negligible at large positive $\l $. This expectation, however, is wrong.
According to (3.19) $\l$ is approaching a $finite$ value as $t \ra \infty $.
Once we ignore the contribution of the potential the damping effect of the
dilaton stops the expansion at a finite $\l$ but for each  finite value of $\l$
the contribution of the potential is still non-vanishing.  The resolution of
this contradiction is that while for $k=0$ the expansion stops at a finite
$\l$,
for $k \not= 0$ it actually  continues towards $\l =\infty$.  In fact, if $\l$
was  approaching a finite value the  asymptotic solution would be a direct
product of the time line  and the $N$ - dimensional maximally
symmetric space. The latter, however, is not a solution (i.e. does not
correspond to  a conformal theory)  if $k$ is non-vanishing (and ``matter"
contributions are absent, cf. (3.3)--(3.7)).

It is easy to see that the  correct large $t$ asymptotics are given by
$$ \l \simeq \l_1 + \ha \  \ln \ t \ , \eq{4.6} $$
$$ \  \vp  \simeq \vp_1 -2bt  - \fourth N \ \ln \ t \
,  \ \ \p  \simeq \p_1 - bt \ +   \fourth N \ \ln  \ t \ ,\eq{4.7} $$
i.e.  the scale factor is slowly growing
while the dilaton is linearly decreasing as in (3.19) to compensate for the
non-vanishing $c$.  We have confirmed this behaviour by the numerical
solution of the system (4.1)--(4.3). As expected, this solution is very
different
from the de Sitter space. We shall discuss how its asymptotics looks like in
the
Einstein frame in Appendix A.\foot {
   One can find analogous solutions also in the
Euclidean case. The system (4.1)--(4.3) and its solutions are transformed
into their Euclidean counterparts by the substitution: $t \ra i\tau\ , \  c \ra
-c \ , \ k \ra -k \ .$ }
Let us  comment on the case when $c > 0\ $ and (or) $\ k > 0$.  If $\dvp (0) <
0
\ , \l (0) > 0 $  the expansion is first slowed down by the dilaton until
$\dvp$
changes sign to a positive one and accelerates the expansion so that $\l$
becomes
infinite in a finite time. As follows from (4.1), (4.3) (cf.(3.17))
$$ \ddot y + c_{\rm eff} (t)\  y = 0 \ , \ \ y = \e{-\vp} \ , \ \
\ c_{eff} = c + N (N-1)k \e{-2 \l} \ . \eq{4.8} $$
For large $\l$ $ \ c_{\rm eff} \simeq c > 0 $ so that
$$ \vp \simeq \vp_0 - \ln \ ( {\rm sin } \ {\sqrt c} \ t )   \  \ ,\eq{4.9} $$
and the solution develops a singularity in a finite time.
If $k > 0 $ and $ c < 0$ there exist solutions
which correspond to an expansion to a maximal radius and subsequent contraction
to zero in a finite time (since $c_{\rm eff}$ in (3.9) is positive for
sufficiently large negative $\l$  $\ \dvp $ changes sign and  the solution
becomes singular in a finite time).

Now it is easy to understand a qualitative
behaviour of the solutions when both  the spatial curvature and
the antisymmetric tensor (or dilaton potential)  contributions
are taken into account. If $k < 0$ the two terms in
$U$ in (3.3) and in (3.12)  have the same sign so that $\ce$ (3.9) is negative
if
$c < 0$.
In the case of the first antisymmetric tensor background (3.2) the
$O(h^2)$ term in $U$ (3.3) is irrelevant at large distances but determines
the behaviour at
small $t$ (assuming $\l (0) $ is large negative). If $a$ starts contracting
it  eventually reflects from the potential wall and expands to infinity.

In the region   $\l < 0$ the potential (3.12) can be approximated  by the first
term. One could expect that the large $t$ (or $\l > 0 $ ) asymptotics of the
corresponding solutions  are determined by the second ``matter" term in $U$.
However, if $\vp$ is decreasing rapidly enough it may dominate over the growth
of $\l$ in the $O(h^2)$ term in (3.12). This is actually  what happens. If one
ignores the spatial curvature term in $U$
the large $t$ limit of the solution (3.20) is $\l \ra \const$ but as we have
discussed above that means that the neglect of the first term in (3.12) is not
justified. The large $t$ limit of the solution is, in fact, dominated by the
curvature term in (3.12), i.e. is again given by (4.6). This conclusion seems
to be valid in the general  case of  $c \leq 0 \ , \ k < 0 $ and the dilaton
potential $V(\p)$ given by a sum of exponentials $\e{ p\p } \ , \ p >0 $
with positive coefficients. In fact, a slow growth of $\l$  and  a rapid
decrease of the dilaton $\p$ with time implies that the dilaton potential term
in $U$  (2.13) will be negligible  at late times.


\vskip .6in

 I would like to acknowledge useful discussions of related issues with
 G. Gibbons and V. Linetsky.
  I am also grateful to Trinity College, Cambridge for
a financial support through a visiting fellowship.

\vskip .3in
\vfill \eject

\appendix{ A } { Relation between cosmological solutions in the string frame
and
the Einstein frame }

We have discussed cosmological solutions using the ``string frame" in which the
action has the form (2.2). Though the \sf and  the \ef  should be physically
equivalent, the metric corresponding to (2.2) is the one which directly appears
in the string action  so it is natural to use it
in order to try to understand  a string theory interpretation of the solutions.
 Also, the form of the
solutions in the \sf is often simpler than that of
their counterparts in the \ef. Below we shall discuss the  relation
between cosmological solutions in the two
frames.\foot{ A similar relation is known in the context of the Brans-Dicke
theory as a relation between Jordan frame and Einstein frame.
It should be emphasized however that the string effective action does $not$
correspond to $\omega = -1$ BD theory because of different couplings of the
dilaton to ``matter" (see also [11]).}

If $D>2$ one can transform the  \sf action (2.2)
$$ S =  \int d^D x \sqG \  \e{-2\p}   \ [  \ R \ + 4 (\pa \p )^2
  - {\hat V}(\p)   - {1 \over 12} H^2_{\l  \mu \nu}-{1 \over 4} F^2_{\mu
\nu } + ...\ ] \ \ ,\eq{A.1} $$
$$ \ \  {\hat V } = V( \p ) - c  \  \ \  , $$
into its \ef form
$$ S_{\E} =  \int d^D x {\sqrt { -g }}   \ [  \ R \ -2p (\pa \p )^2
  - { \e{2p\p } }  { \hat V}(\p)  $$  $$  - {1 \over 12}\e{-4p \p }
  H^2_{\l  \mu \nu}-{1\over 4} \e{ -2p \p } F^2_{\mu \nu} + ...\ ] \
\ ,\eq{A.2} $$
which depends on the new metric $g_{\mu \nu }$
$$  g_{\mu \nu }= \e{-2 p\p } G_{\mu \nu }\ \ , \ \ p\equiv {2\over D-2} \ .
\eq{A.3} $$
Given a cosmological solution (2.1) in the \sf theory (A.1)
$$ds^2 = \Gmn dx^{\mu } dx^{\nu }= - dt^2 +  \e{2\l (t) } \ d\Omega^2 \ \ ,
\ \ \p = \p (t) \ \ , \eq{A.4} $$
 we can find the corresponding solution in the \ef theory (A.2)
$$ds^2_{\E} = g_{\mu \nu } dx^{\mu } dx^{\nu }= \e{-2p\p (t) }
(- dt^2 +  \e{2\l (t) } \ d\Omega^2 \ ) \ \ ,  \  \eq{A.5} $$
i.e.
$$ ds^2_{\E} = - d\t^2 +  \e{2\le (\t) } \ d\Omega^2 \ \ ,  \
\ \  d\t = dt \ \e{- p \p (t) } \ \ , \eq{A.6} $$
$$ \le  \equiv \l - p \p  \ \ . \eq{A.7} $$
Using (2.9), i.e.
$$ \p = \ha (\vp + N \l )  \ , \ \ \ \ N=D-1 \ \ ,  \eq{A.8} $$
we can rewrite (A.7) in the form
$$\le =\l - \fourth p (\vp + N \l ) = - {1\over N-1} ( \l + \vp ) \  \ \ .
\eq{A.9} $$
If $\p = \const $ the solutions look of course the same in the two frames. Once
$\p \not= \const$  the relations (A.6)--(A.8) suggest that a correspondence
between the behaviour of the functions $\l (t) \ , \p (t) $ and $\le (\t ) \ ,
\p (\t ) $  may be non-trivial.  Suppose that $\l (t) $ is growing while $\vp
(t) $ is decreasing rapidly enough so that $\p (t) $ is also decreasing and is
monotonic.\foot{Note that our assumption that $\dvp < 0 $ is only a
necessary but not   sufficient condition for a decrease of the dilaton $\p (t)
$. }
Then according to (A.6)  the \ef time $\t $ is a monotonically growing function
 of $t$. As a consequence, $\p (\t) $ is still decreasing  while $\l (\t ) $ is
increasing so that  (A.7) implies that $\le (\t ) $  is also increasing.  In
this case an expansion in the \sf corresponds to an expansion in the \ef.
However, the rates of expansions may be quite different. As follows from (A.7),
(A.9) $$ \le' = \e{p \p } ( \dl - p \dot \p ) =
{1\over N-1} {\e{p \p } }(  - \dvp - {\dl } )  \ \ , \ \ \le' \equiv {  d\le
\over d\t  } \ \ ,\eq{A.10} $$
so that  $\le'$ depends strongly on the behaviour of the dilaton.

To illustrate the above general remarks let us consider the following example
$$ \l = \l_0 + q \  \ln \ t \ , \ \  \p  = \p_0 - bt \ ,
\  \ \ \vp \simeq \vp_0 - 2bt \  , \ \  \ b >0 \ \ . \eq{A.11} $$
Then
$$ \t = \t_0 + m^{-1}{\rm e}^{2bt \over N-1} \ ,
\ \  m = {2 b\over N-1  } \ \e{{2\p_0 \over N-1} } \ \ , $$
$$
\ \ \p = \p_0 - \ha (N-1) \ {\ln  \ }m (\t -\t_0 ) \
, \ \eq{A.12} $$
$$ \le = {\le }_0 +  \ {\ln  \ } m(\t -\t_0 )  + q \ \ln \ {\ln  \ }
m(\t - \t_0) \ \ . \eq{A.13} $$ While $\p (\t)
$ is decreasing much slower than $\p (t) $ ,  $ \ \le (\t) $ is  still growing
logarithmically with the coefficient of the leading logarithm being
$universal$, i.e. independent of  $b$
in the dilaton $\vp (t) $ or   $q$ in $\l (t) $. As a result, the static metric
( $q=0$) in the \sf  corresponds to the linear expansion $ a_{\rm E } (\t )
\sim  \t $ in the \ef [5,10,11]. The large $\t $ asymptotics in  $D=4$ is
$$ \le  \simeq +  \ {\ln  \ } \t  \ \ , \ \
\ \p \simeq -   \ {\ln  \ } \t  \  \  . $$
Solutions with such asymptotics were discussed in [5,6,19,20].

Another useful example is  $$ \l = \l_0 + q \  \ln \ t \ , \ \  \p  = \p_0  - s
\ {\ln }\ t  \ ,  \ \  \ \vp  = \vp_0  - (2s +  N q) \ {\ln }\ t  \ \ .
\eq{A.14} $$ Assuming $ sp \not= 1$ we get
$$\t = \t_0 +  m^{-1} t^{sp-1}  \ \ , \ \ m = \e{ p\p_0 } (sp-1) \ \ , $$ $$
\ \ \p = \p_0 - r\ {\ln  \ }|m (\t -\t_0 )| \ \ ,
\ \ \le = { \le }_0 + l \ {\ln  \ }| m(\t - \t_0 )|
\ \ , \eq{A.15} $$
$$\ \ r ={s \over sp-1 } \ , \ \ l = { q + sp \over sp-1} \ \ . $$
If $sp=1$
$$ \t = \t_0 +  m^{-1} \ {\ln }\t \ \ , \ \ m = \e{ p\p_0 } \ ,
\p = \p_0 - s m (\t -\t_0 ) \ , \ \ \le = {\le }_0 + q m(\t - \t_0 ) \
\ ,   $$
i.e. we get the exponential inflation of $a_\E = \e{ \le } .$
If $sp <1$ we obtain  (A.15) with negative $ m,r,l$ and $\t$ asymptotically
approaching $\t_0$ from below as $ t \ra \infty .$

We are now able to  discuss the \ef form of the solutions found in sections 3
and
4. The singular solutions with $\ce>0$ remain singular in the \ef so we shall
consider the case of $\ce\leq 0$.  The $k>0$ solutions  corresponding to the
potential (3.3)  asymptotically approach the ``static" solution (3.7). Using
(A.11)--(A.13) we conclude that their large $\t$  behaviour in the \ef is
 represented by
$$  \le \simeq {\le }_1 +  \ {\ln  \ } \t  \ \ ,
\ \ \p \simeq \p_1 - \ha (N-1) \ {\ln  \ } \t  \ \ ,  \   \eq{A.16} $$
 i.e. is  the linear expansion with a logarithmically decreasing dilaton.

The exact $k=0$ solution (3.20), (3.18)  found in the case of the potential
(3.12)  looks very complicated in the \ef but its large $\t$ asymptotics is
easily found from (3.21). It is again given by (A.16).
The  small $t$ limit of the $c=0$ solution (3.22)
$$ \l = \l_0 + {1 \over \sn }  \  \ln \ t \ \  ,   \ \  \ \vp  = \vp_0  - \
{\ln \ } t  \ \  , \ \   \p  = \p_0  - (1- \sn ) \ {\ln \ } t  \   \eq{A.17} $$
can be translated into the \ef  using (A.14), (A.15) (note that here $s<0$).
We find (A.15) with $l>0\ , \ r>0 $, i.e. the scale factor is growing while the
dilaton is decreasing with $\t$.

Let us now turn to the solutions in the case  of  the ``pure curvature"
potential (4.1)--(4.3). The special  solution (4.5) (which is valid in the
small $t$ region) looks
the same in the \ef since $\p = \const $. The large $t$ asymptotics (4.6),
(4.7)
is of the type (A.11) with $q=\ha $ so that the corresponding $ \p (\t
) \ , \ \le (\t) $ are given by (A.12), (A.13), i.e. (A.16).

For completeness let us record  the \ef analog of the system of cosmological
equations (2.18)--(2.20)
$$ N(N-1) { \le' }^2 -  {4\over N-1} \p'^2 = 2\ \CU  \ \ , \eq{A.18} $$
$$ (N-1) \le'' + {4\over N-1} \p'^2 = {1 \over  N} {\pa \CU \over \pa \le} \ \
,
\eq{A.19} $$
$$  \p'' + N \le'\p' =- \fourth (N-1) {\pa \CU \over \pa \p }  \ \ ,  \eq{A.20}
$$
$$ \CU (\le , \p )  \equiv  \e{{4\p \over N-1}  } \ [ \ U (\l , \vp ) - \ha c \
]  \ \ .  \eq{A.21} $$
Here  primes denote derivatives over $\t$ and $U$ is given by (2.13). The
expression for $\CU$ can be derived directly from (A.2). As a simple example
let us consider the case when $U$ contains only the curvature term and
$c=0$.  Then
$$\CU = - \ha N(N-1) \e{ -2\le } \ \ , \ \  \p'' + N \le'\p' =0  \ \ ,
 \eq{A.22} $$
i.e.
$$ \p' = - q \e{-N \le } \ \ , \ \ { \le' }^2 = -k \e{-2\le } + {4q^2\over
N(N-1)^2 } \e{-2N \le } \ \ . \eq{A.23} $$
If $k>0$ there exists a maximal radius of expansion  while for $k<0$ the
expansion  continues to $\le = \infty $. The case of $c\not= 0 $ is obviously
easier to analyse in the \sf.

\appendix{B}{ Correspondence between the $N=1$ solution (3.18), (3.20)
and the  $D=2$ charged black hole background }

If $N=1$ the metric, the dilaton $\p$  and the gauge field corresponding to
(3.20), (3.18)  are given by
$$ ds^2 = -dt^2 + \e{2\l(t)} dx^2 \  ,\ \ $$ $$
\l = \l_0  - \  {\ln} [\  A^{-1}{({\th }bt)}^{-1}
+ A \ {{\th }bt} \ ]\ , \eq{B.1} $$
$$ \p = \ha ( \vp  +\l) = \p_0 -\ha{\ln }( A^{-1} \ {\rm cosh}^2 \ bt
+ A \ {\rm sinh}^2\ bt) \  \ , \eq{B.2} $$
$$ F_{\mu \nu } = h \e{2 \p} \ep_{\mu \nu} \ \ , \ \  A= {h\over 4 {\sqrt 2 }
b}
\e{2\p_0} \ . \eq{B.3} $$
This is to be compared with the  $D=2$ charged black hole metric of ref. [36]
$$ ds^2 = -g dy^2 + g^{-1} dr^2 \ , \ \p=\p_0 - \ha Q r \ \ ,
\ \ c= Q^2 \ , \eq{B.4} $$
$$  g= 1- 2m\e{-Qr} + q^2 \e{-2Qr} \ \  . \eq{B.5} $$
The metric (B.5) has two horizons and a singularity at $r=-\infty$.
Introducing the new time parameter $z$ one can represent (B.1), (B.2)
in the  form
$$ ds^2 = -f^{-1} dz^2 + f dx^2 \ ,\ \ \p=\p_0 - b z  \ , \ \ c = -4b^2
\ , \eq{B.6} $$
$$ \ \  f = (1 + A^2 \e{-2bz } ) ( 1-\e{-2bz } ) \  . \eq{B.7} $$
The relation between  (B.4), (B.5) and (B.6), (B.7) is established by the
following identification
$$ r=iz+r_0\ , \ \ y = ix \ , \ \ g=f \ , \ \ Q=-2ib \ , \ \ 2m\e{-Qr_0 }
=1-A^2
\ ,  \ \ q \e{-Qr_0 } = iA  \ . \eq{B.8} $$
As in the case of the zero charge black hole [22,17] the cosmological solution
corresponds to the region between the horizons and singularity.

\appendix{ C } {
Time--dependent solutions interpolating between ``static"  conformal
backgrounds}

The system of cosmological equations (2.18)--(2.19) has another interesting
interpretation.
Suppose the potential $U (\l ) $ has a maximum and a minimum. The two
extremal points are solutions of (2.18)--(2.19) (for appropriate
 signs of $c$ and $U$ and a linear dilaton). If we start from the maximum we
shall end the evolution at the minimum, i.e.  $\l(t)$  will be  interpolating
between the two ``static" solutions.  This is similar to  the situation one
encounters  in the case of a conformal theory perturbed by a nearly
marginal operator. In this context  $\l$ will be interpreted as  a coupling
associated with a nearly marginal operator. There are two conformal points: the
original one ($\l=0$) and the nearby one. Embedding into string theory, i.e.
introducing time--dependent dilaton and demanding the vanishing of the total
central charge we shall get a system of equations for $\vp (t)$ and $\l (t)$
(see also [39,41]). Below we shall discuss a relation between this system and
the
standard renormalisation group equation.

A simple way of deriving this system is the following. Consider the effective
action for the metric, dilaton and tachyon couplings (cf.(2.2); $\a'=1 $)
$$ S =  \int d^D x \sqG \  \e{-2\p}   \ [ \ c + \ R \ + 4 (\pa \p )^2
  - \fourth (\pa T )^2  + T^2 - {1\over 6} T^3 + ... \ ] \ . \ \eq{C.1} $$
Suppose now that the dependence of $T$ on the space coordinates is such that it
is nearly ``on-shell" in $N$ space dimensions, i.e. an effective mass of a
``reduced" field is
small. Assuming that the metric (2.1) is spatially flat and isotropic
and that the dilaton  depends only on $t$ let us  replace $T$ by a
function of time only  $T(t,x) \ra 2 f (t)$.\foot{For example, in $N=1$ case
this can be done
 by setting $T \sim f(t) \cos px $ (with a properly chosen
$p$) and integrating over $x$ (more generally, $ T =\sum_{n} f_n(t)\  \e{ i
p_n\cdot x } $ and integration over $x^a$ imposes ``momentum conservation";
cf. [42,43]). } As a result,  we find the following ``dimensionally reduced"
action (cf. (2.10), (2.21))  $$ S= \int dt\  \e{-\vp} [ \ c + N\dl^2
 -\dvp^2 \   + {\dot f } ^2   - 2U ( f) \ ] \ \ . \eq{C.2} $$
$$ U = - 2\m f^2 +  {2 \over 3 }g f^3 + ...\  , \ \ \  \ \ 0<\mu \ll 1 \  \ ,
\eq{C.3} $$ where $g$ is an effective coupling. The corresponding system of
evolution equations is similar to (2.18)--(2.19) (except that now $U$ does not
depend on $\vp$ and $\l$ )
$$ c- N\dls -{\dot f }^2 + \dvps  = 2U  \ \ , \eq{C.4} $$
$$  \ddot f - \dvp \dot f = - U' \  \ , \ \ \ \ U'\equiv  {\pa U \over \pa f} \
\
, \eq{C.5} $$
$$  \ddvp - N \dls -  {\dot f } ^2 =0 \ \ ,  \eq{C.6} $$
$$ \ddl - \dvp \dl =0 \ \ . \eq{C.7} $$
 $f$ appears in the l.h.s. of equations of this system in the same way as $\l$
does (or as one of the ``moduli" $\l_i$ in (2.28)--(2.30)).
 $\l$ is easily eliminated by solving (C.7)
\foot{ The result of solving (C.7) and substituting $\l$ (i.e. $ \dl = h \e{\vp
}
$) in the remaining equations is equivalent to adding the term $ \ha Nh^2
\e{2\vp
} $ to the potential $U$   in (C.4) and including its derivative over $\vp$ in
(C.6) (cf. (2.20)). Since $\vp$ is decreasing with time  the effect of this
term is not important. In general $\l$ will be slowly changing with time
asymptotically approaching  a constant value. }
but let us consider for simplicity
the  trivial solution $\l = \const$ (flat metric). In this case there is no
difference in the behaviour of the original dilaton $\p$ and the ``shifted" one
$\vp$ (2.9). The resulting system of equations for  $\vp(t)$ and $f(t)$ is
$identical$ to (2.18)--(2.20) with $N=1,\  U=U(\l) , \ \l = f $
 $$ c -{\dot f }^2 + \dvps  = 2U  \ \ , \eq{C.8} $$
$$  \ddot f - \dvp \dot f = - U' \  \ , \ \eq{C.9}
$$  $$  \ddvp - {\dot f }^2 =0 \ \ .  \eq{C.10} $$
For the potential (C.3)  and $c \leq 0$ this system has  two ``static"
solutions corresponding to the extremal points (the maximum and the minimum)
of
$U$
 $$ f=f_0 =0\ \ , \ \ \  \vp = \vp_0 - q_0  t  \ \ , \ \ \ \ \  q_0 > 0 \ \ , \
\   c+ q_0^2 =0 \ \ , \eq{C.11} $$
$$  f = f_1 = {2 \m \over g} \ \ , \ \ \
\vp = \vp_1 - q_1  t  \ \ , \ \  \eq{C.12} $$
 $$  c+ q_1^2 = 2U(f_1) = -{16\m^3 \over 3 g } \ \ , \ \ \ \
 q_1^2 - q_0^2 =  -{16\m^3 \over 3 g } \ \ . \eq{C.13} $$
Following the discussion in Sect. 2 it is easy to analyse the behaviour of
time--dependent solutions of (C.8)--(C.10).  Since the maximum of the potential
is an unstable point the solution with (C.11) and $\dot f = \epsilon >0 $ as
the
initial conditions   will asymptotically approach the (stable) minimum point
(C.10). The dilaton is decreasing with time (with $\dvp$ increasing from $-q_0$
to $-q_1$) thus providing a damping force which eliminates oscillations near
the
minimum. Note that the effective central charge of the  theory   ($\sim -c
-\dvps$ ) is $decreasing$ with time (in our notation $c$ has the sign opposite
to that of the standard central charge $\CC\ $ , $\  c= --{2 \over 3 }\CC $).

This is of course reminiscent of the well--known picture of the RG
interpolation between nearby conformal theories [44].
Let $\g^i$ be a set of couplings which parametrize
a perturbation of a conformal theory by local operators $\CO_i$ of dimensions
$2-w_i$. Then the leading terms in the corresponding renormalisation group
$\beta$ - function are expressed in terms of $w_i$ and the operator product
coefficients $C_{ijk}$ (in normal coordinates Zamolodchikov's metric is trivial
 to the leading order)
$$ \beta^i = {d\g^i  \over d \t } = w_i \g^i -\pi \sum_{j,k} C_{ijk} \g^j \g^k
 + O(\g^3)  \ \ , \eq{C.14} $$
where $\t$ is the RG ``time" parameter.
If one of the operators (say $\CO_1$) is nearly marginal (i.e $ w_1 \ll 1 $)
 it  drives the system to another conformal point corresponding to  the
non-trivial zero of  (C.14)
$$  \g^1 = {w_1 \over \pi C_{111} } + O (w_1^2) \ \ , \ \ \ \g^i = O
(w_1^2) \ \ ,  \ \ i=2,3,... \ \ , \ \eq{C.15} $$
with the difference of central charges being [44]
$$ \CC (1) -\CC (0) = - {w_1^3 \over \pi C_{111} } + O (w_1^4) \ \ . \ \
\eq{C.16} $$ These relations  are in direct correspondence with  (C.11)--(C.13)
with $f= \g^1 \ , $ $\ \mu \sim w_1 , \  g\sim C_{111}. $

Ignoring the effects of other  couplings the RG equation for  $f(t) = \g^1
(\t) $ where  $  t=q\t  $ is a rescaled ``time" can be written in the form
$$ q \dot f = - U' \ \ , \ \ \ \ \dot f = {df\over dt }
\eq{C.17} $$  $$ U = - 2\m f^2 +  {2 \over 3 }g f^3 + ...\ \ . \ \eq{C.18} $$
Let us compare (C.8)--(C.10), (C.3)  with (C.17), (C.18).  Their static
solutions ($f= 0,\ f_1 $) are in correspondence-- the only difference is
that (C.8)--(C.10) contains also the dilaton which ensures that the total
central
charge is zero. The time--dependent solutions $f(t)$ of
(C.8)--(C.10) and of (C.17) thus interpolate between the same asymptotic
conformal points. However, they are obviously $different$ as functions of $t$.

The system (C.8)--(C.10) can be considered as a ``string generalisation" of the
standard RG equation (C.17): to satisfy the Weyl invariance condition in
``$N+1$ dimensions" (i.e. the vanishing of the ``N+1- dimensional"
$\b$-functions
and of the total central charge) we need to introduce a time-dependent dilaton
with its own  equation of motion and to add the second derivative term to
(C.17).
One can try to establish a correspondence between (C.8)--(C.10) and (C.17)
 in a ``semiclassical" limit of large negative $c$ [40]: setting $\dvp = -q +O
(q^{-1})  \ , \ \ q \gg
1\ $ and  $t = q \t $ we get from (C.8)--(C.10)
 $$ c  + q^2  =   2U + O(q^{-2}) \ \ , \eq{C.19} $$
$$   {df\over d\t }  = - U' + O(q^{-2}) \  \ , \ \eq{C.20}
$$  $$  q^{-2}({df\over d\t })^2 = O(q^{-2})  \ \ .  \eq{C.21} $$
Eq.(C.19) reduces to (C.17) but the zero central charge equation (C.19) is
$not$
satisfied in general. It is only if $U$ is very small on the
solution (which is true only for particular potentials $U$, e.g.
$U=- b^2 f^n $) that one  can claim an agreement between (C.17) and
(C.19)--(C.21) (compare [40]).

It is interesting to note that there exists a potential
$U$ for which any solution of (C.17) is also a solution of (C.8)--(C.10).
Substituting (C.17) into (C.9), (C.10) one finds that $U$ should satisfy
 $$U'''=U'\ \ , \ \ \dvp = -q - q^{-1}U'' \ \ . \eq{C.22} $$
As a result,
$$U = U_0 +  A  \ch{ f }  + B \sh{ f } \ \ ,  \ \eq{C.23} $$ $$ \ \ c + q^2 +
A^2
- B^2 = 2U_0 \ \ , \eq{C.24} $$
where (C.24) follows from (C.8). It is because of the
``dissipative" nature of the dilaton coupling that solutions of the second
order system (C.8)--(C.10) can in principle be in correspondence with the
solutions of the first order RG equation (C.17).

\vfill\eject

\centerline{\bf References}
\bigskip

\item {[1]} D. Bailin, A. Love and D. Wong, \pl B165(1986)409 ;

 D. Boulware and S. Deser, \pl B175(1986)409 ;

   Y.S. Wu and Z. Wang, \prl 57(1986)1978 ;

   J.T. Wheeler, \np B268(1986)737 ;

 A. Henriques, \np B277(1986)621 ;

   A. Henriques and R. Moorhouse, \pl B197(1987)353 ;

   D. Lorentz--Petzold, \pl B197(1987)71 ;

   K. Maeda, \mpl A3(1988)243 .

\item {[2]} I. Antoniadis and C. Kounnas, \np B284(1987)71 ;

S. Kalara, C. Kounnas and K.A. Olive, \pl B215(1988)265 .

\item {[3]} G. Gibbons and P. Townsend, \np B282(1987)610 ;

 G. Gibbons and K. Maeda, \np B298(1988)741 .

\item {[4]}  R. Myers, \pl B199(1987)371 .

\item {[5]} I. Antoniadis, C. Bachas, J. Ellis and D.V. Nanopoulos,

\pl B211(1988)393; \np B328(1989)115 .

\item {[6]}S. Kalara and K.A. Olive, \pl B218(1989)148 .

\item {[7]} A. Liddle, R. Moorhouse and  A. Henriques, \np B311(1988)719 ;

N. Stewart,  Class. Quant. Grav. 8(1988)1701 ;

\item {[8]}R. Brandenberger and C. Vafa, \np B316(1988) 391 ;

 H. Nishimura and M. Tabuse, \mpl A2(1987)299 ;

 J. Kripfganz and H. Perlt, Class. Quant. Grav. 5(1988)453 ;

M. Hellmund and J. Kripfganz, \pl B241(1990)211 .

\item {[9]}M. Mueller, \np B337(1990)37 .

\item {[10]} N. Sanchez and G. Veneziano, \np B333(1990)253 .

\item {[11]}B.A. Campbell, A. Linde and K.A. Olive, \np B355(1991)146 .

\item {[12]} I. Antoniadis, C. Bachas, J. Ellis and D.V. Nanopoulos,

       \pl B257(1991)278 .

\item {[13]}J.A. Kasas, J. Garcia--Bellido and M. Quiros, \np B361(1991)713 ;

 J. Garcia--Bellido and M. Quiros, \np B368(1992)463 ;

 M.C. Bento, O. Bertolami and P.M. Sa, \pl B262(1991)11 .

\item {[14]} A.A. Tseytlin, \mpl A6(1991)1721 .

\item {[15]} G. Veneziano, \pl B265(1991)287 .

\item {[16]}  A.A. Tseytlin, in: Proc. of the First International A.D. Sakharov
Conference on Physics, Moscow  27 - 30 May 1991, ed.  L.V. Keldysh et al.,
Nova Science Publ., Commack, N.Y., 1991 .

\item {[17]}A.A. Tseytlin and C. Vafa, \np B372(1992)443 .

\item {[18]}A.A. Tseytlin, preprint DAMTP-37-1991; Class. Quant. Grav. 9(1992)1
{}.

\item {[19]}B.A. Campbell, N. Kaloper and K.A. Olive, Univ. of Minnesota
preprint
  UMN-TH-1006/91.

\item {[20]} N. Kaloper and K.A. Olive, Univ. of Minnesota preprint
UMN-TH-1011/91.

\item {[21]}I.Bars and D. Nemeschanski, \np B348(1991)89 ;

I. Bars, U. Southern California preprint USC-91/HEP-B4 ;

E.S. Fradkin and V.Ya. Linetsky, \pl B261(1991)26  .

\item {[22]} E. Witten, \pr D44(1991)314 ;

R. Dijgraaf, H. Verlinde and E. Verlinde, Princeton preprint PUPT-1252/91.

\item {[23]} M. Crescimanno, Berkeley preprint LBL-30947(1991) .

\item {[24]}I. Bars and K. Sfetsos, U. Southern California preprints
USC-91/HEP-B5 ;  USC-91/HEP-B6 .

\item {[25]} E.S. Fradkin and V.Ya. Linetsky, Harvard preprint
 HUTP-91/A044 (1991) .

\item {[26]} A.H. Chamseddine, Zurich U.  preprint ZU-TH-31-1991 .

\item {[27]}P. Ginsparg and F. Quevedo, Los Alamos preprint LA-UR-92-640 .

\item {[28]}P. Horava, Chicago U. preprint EFI-91-57 ;

D. Gershon, Tel-Aviv U. preprint TAUP-1937-91 .

  \item {[29]}  S. Elitzur, A. Forge and E. Rabinovici, Nucl. Phys. B359 (1991)
581;

G. Mandal, A. Sengupta and S. Wadia, Mod. Phys. Lett. A6(1991)1685.

\item {[30]} E. Witten, Commun. Math. Phys. 92(1984)455 .

\item {[31]} J. Scherk and J.H. Schwarz, Nucl. Phys. B81(1974)118 .

\item {[32]} E.S. Fradkin and A.A. Tseytlin, \pl B158(1985)316 ;

 C.G. Callan, D. Friedan, E. Martinec and M.J. Perry, \np B262(1985)593 .

\item {[33]} P.G.O. Freund, \np B209(1982)146 ;

P.G.O. Freund and M. Rubin, \pl B97(1980)233 .

\item {[34]}M. Gleiser, S. Rajpoot and J.G. Taylor, \pr D30(1984)756; Ann. of
Phys. 160(1985)299 ;

 D. Bailin, A. Love and J. Stein--Schabes, \np B253(1985)387 ;

 F.S. Acceta, M. Gleiser, R. Holman and E.W. Kolb, \np B276(1986)501 ;

 R. Holman, E.W. Kolb, S.L. Vadas and Y. Wang, \pr D43(1991)995 .

\item {[35]} S. Randjbar-Daemi, A.Salam and J. Strathdee, \np B214(1983) 491 ;

E. Sezgin and A. Salam, \pl B147(1984)47 ;

K. Maeda and H. Nishino, \pl B158(1985)381 ;

Y. Okada, \np B264(1986)197 ;

J. Halliwell, \np B286(1987)729 ;

A. Linde and M. Zelnikov, Phys. Lett. B215(1988)59 .

\item {[36]} M.D. McGuigan, C.R. Nappi and S.A. Yost,
Princeton preprint IASSNS-HEP-91/57.

\item {[37]}  I. Antoniadis, C. Bachas and A. Sagnotti, \pl
B235(1990)255 .

\item {[38]} V.Ya. Linetsky, unpublished.

\item {[39]} S.Das, A.Dhar and S. Wadia, \mpl A5(1990)799 ;

A. Sen, \pl B252(1990)566 .

\item {[40]} A. Cooper, L. Susskind and L. Thorlacius, \np B363(1991)132 ;

A. Polyakov, Princeton U. preprint PUPT-1289 (1991).

\item {[41]} S. Mukherji, Tata Inst. preprint TIFR/TH/92-11 ;

\item {[42]}  M.T. Grisaru, A. Lerda, S. Penati and D.Zanon, \pl B234(1990)88 ;
\np B346(1990)264 .

\item {[43]} A.A. Tseytlin, \pl B241(1990)233 ; \pl B243(1990)465(E) .

\item {[44]} A.B. Zamolodchikov, Sov. J. Nucl. Phys. 46(1987)1090 ; JETP Lett.
43(1986)731 ;

J.Cardy and C.Ludwig, \np B285(1987)687 .

 \vfill
\eject
\bye